\documentclass{aa}
\usepackage{graphicx}
\usepackage{natbib}
\bibpunct{(}{)}{;}{a}{}{,}
\begin{document}
   \title{Weak lensing measurements of dark matter halos of galaxies from
   COMBO-17}

   \author{M. Kleinheinrich\inst{1,2}
          \and P. Schneider\inst{2}
          \and H.-W. Rix\inst{1}
          \and T. Erben\inst{2}
          \and C. Wolf\inst{3} 
          \and M. Schirmer\inst{2}
          \and K. Meisenheimer\inst{1}
          \and A. Borch\inst{1}
          \and S. Dye\inst{4}
          \and Z. Kovacs\inst{1}
          \and L. Wisotzki\inst{5}
          }

   \offprints{M. Kleinheinrich,\\ \email{martina@mpia.de}}

   \institute{Max-Planck-Institut f\"ur Astronomie, K\"onigstuhl 17, 
              D-69117 Heidelberg, Germany 
              \and 
              Institut f\"ur Astrophysik und Extraterrestrische Forschung, 
              Universit\"at Bonn, Auf dem H\"ugel 71, 53121 Bonn, Germany
              \and
              Department of Physics, Denys Wilkinson Bldg., University of 
              Oxford, Keble Road, Oxford, OX1 3RH, U.K.
              \and 
              School of Physics and Astronomy, Cardiff University, 5 The 
              Parade, Cardiff, CF24 3YB, U.K.
              \and
              Astrophysikalisches Institut Potsdam, An der Sternwarte 16, 
              D-14482 Potsdam, Germany
             }

   \date{Received / Accepted}

   \abstract{We present a measurement of mass estimates for dark
   matter halos around galaxies from the COMBO-17 survey using weak
   gravitational lensing. COMBO-17 is particularly useful for this
   kind of investigation because it covers observations in 17 optical
   filters from which accurate photometric redshifts and spectral
   classification for objects with $R<24$ are derived. This allows us
   to select lens and source galaxies from their redshifts and to thus
   avoid any uncertainties from estimates of the source redshift
   distribution. We study galaxy lenses at redshifts
   $z_\mathrm{d}=0.2-0.7$ by fitting the normalization of either
   singular isothermal spheres (SIS) or Navarro-Frenk-White (NFW)
   profiles to the whole lens sample; we then consider halos around
   blue and red subsamples separately. We also constrain the scaling
   of halo mass with light. For the NFW model, we find virial masses
   $M_\mathrm{vir}^*=3.9^{+3.3}_{-2.4}\times 10^{11}h^{-1}M_{\sun}$
   (1-$\sigma$) for blue and $M_\mathrm{vir}^*=7.1^{+7.1}_{-3.8}\times
   10^{11}h^{-1}M_{\sun}$ for red galaxies of
   $L_\star=10^{10}h^{-2}L_{\sun}$, respectively. The derived
   mass-to-light scaling relations suggest that the mass-to-light
   ratio might decrease with increasing luminosity for blue
   galaxies but increase with increasing luminosity for red
   galaxies. However, these differences between blue and red galaxies
   are only marginally significant and both subsamples are consistent
   with having the same mass-to-light ratio at all luminosities. Finally,
   we compare our results to those obtained from the Red-Sequence
   Cluster Survey (RCS) and the Sloan Digital Sky Survey
   (SDSS). Taking differences in the actual modelling into account, we
   find very good agreement with these surveys.

   }

   \maketitle
%
%________________________________________________________________

\section{Introduction}
\label{sect: introduction}

Dark matter is the dominant mass component in the universe and also the
major constituent of cosmological structures like galaxy clusters or
galaxies. Therefore, dark matter plays a fundamental role in the formation and
evolution of these structures. Our understanding of structure formation is
thus limited by our ability to map the dark matter distribution in these
objects. In this paper, we will investigate the dark matter distribution
around galaxies.

Several methods have been applied to measure masses of galaxies: Rotation
curves of spiral galaxies provided the first evidence for dark matter in
galaxies \citep[e.g.][]{sofue2001}. In elliptical galaxies, the dynamics of
e.g.\ the stellar population itself, globular clusters or planetary nebulae
can be used \citep[e.g.][]{danziger1997}. However, these methods can only be
applied over radii where luminous tracers are available (a few tens of kpc)
and suffer from typical problems of dynamical studies, e.g.\ the unknown
degree of anisotropy \citep{rix1997}. In a few cases it is possible to measure
masses from X-rays or, within the Einstein radius, from strong lensing
\citep[e.g.][]{kochanek2004}. 

Only two methods are currently in use to probe dark matter halos of
galaxies at scales of about $100 h^{-1}~\mathrm{kpc}$ or larger: weak
gravitational lensing and the dynamics of satellite galaxies. At these
scales, the baryonic contribution to the mass is negligible, so that
basically just the dark matter around galaxies is probed. In the
beginning, satellite dynamics was only applied to isolated spiral
galaxies \citep{zaritsky1993,zaritsky1994,zaritsky1997}. 
Later, early-type galaxies were investigated
\citep{mckay2002,prada2003,brainerd2003,conroy2004}, and now galaxies
in denser environments and at much fainter magnitudes are also studied
\citep{vandenbosch2004}. Weak gravitational lensing has become a
standard tool in recent years. Its main advantage over
satellite dynamics it that no assumptions on the dynamical state of the
galaxies under consideration have to be made.  Galaxy-galaxy lensing
is the technique which uses the image distortions of background
galaxies to study the mass distribution in foreground galaxies.
Galaxy-galaxy lensing and satellite dynamics are independent methods
and it is very desirable to have both methods available for comparison
of results. Weak gravitational lensing is similar to the use of
satellite dynamics in the sense that only statistical investigations
are possible due to the weakness of the gravitational shear and the
small number of satellites per primary galaxy. The most
recent result from both techniques are summarized in
\citet{brainerd2004}.

Galaxy-galaxy lensing was measured for the first time by
\citet{brainerd1996}. In early work, people concentrated on isothermal
models for describing the lenses 
\citep{brainerd1996,griffiths1996,dellantonio1996,hudson1998,fischer2000,smith2001,wilson2001,hoekstra2003}.
The best-constrained parameter was the effective halo velocity dispersion for
$\sim L_\star$ galaxies, assuming a scaling relation between the velocity
dispersion and the luminosity according to the Tully-Fisher and Faber-Jackson
relations. The Tully-Fisher index $\eta$ in this relation (see Sect.\
\ref{sect:sis}) was typically assumed according to measurements from the
central parts of the galaxies. Only \citet{hudson1998} were able to
put at least a lower limit on $\eta$. More generally, the galaxy-mass
correlation function was later investigated
\citep{mckay2001,sheldon2004} and the Navarro-Frenk-White (NFW)
profile was considered
\citep{seljak2002,guzik2002,hoekstra2004}. 
The Sloan Digital Sky Survey (SDSS) has turned out to be extremely powerful
for galaxy-galaxy lensing studies \citep{fischer2000,mckay2001,seljak2002,guzik2002,sheldon2004,seljak2004}. Most of its success is due to the fact
that the SDSS provides a large sample of lens galaxies with measured
spectra. The spectra provide very accurate redshifts and classification of
the lens galaxies. Therefore, it is possible to measure dark matter halos as
function of luminosity, spectral type or environment of the galaxies. However,
the SDSS is a very shallow survey and is thus only able to measure lens
galaxies around $z_\mathrm{d}\approx 0.1$. The question of halo properties at
higher redshift and of evolution cannot be addressed. Therefore, substantial
effort goes into measurements of galaxy-galaxy lensing at higher redshift with
at least rough redshift estimates for the lenses from e.g.\ photometric 
redshifts. Currently available are the Hubble Deep Fields which are far
too small to provide statistically clean samples free from cosmic
variance. \citet{wilson2001} addressed the question of evolution, but only for
early-type galaxies. \citet{hoekstra2003} had redshifts available for part of
their lenses, but not yet enough to split the lenses into subsamples and
to study their properties separately. \citet{hoekstra2004} use the larger
Red-Sequence Cluster Survey (RCS) with deep observations. No redshift or
colour information is available, yet, so that again only properties averaged
over all classes of galaxies can be investigated.

Here, we will use the COMBO-17 survey which is a deep survey, that provides
accurate photometric redshifts and spectral classification from observation in
a total of 17 filters \citep{wolf2001,wolf2003,wolf2004}. This data set allows
us to probe lens galaxies at higher redshift ($z_\mathrm{d}=0.2-0.7$), to
derive the relation between luminosity and velocity dispersion (or mass)
instead of assuming it and to measure dark matter halos for blue and red
galaxies separately. In addition to the singular isothermal sphere (SIS) we
will also apply the NFW profile. 

This paper is organized as follows: in Sect.\ \ref{sect:data} we briefly
describe the data set and in Sect.\ \ref{sect:method} our method of measuring
galaxy-galaxy lensing. Section \ref{sect:sis} gives our results from the SIS
model and Sect.\ \ref{sect:NFW} those from the NFW profile. In Sect.\
\ref{sect:clusters} we investigate how our measurements are affected by the
presence of large foreground clusters in one of the survey fields. In Sect.\
\ref{sect:comparison} we will compare our result to those from the RCS and the
SDSS. We close with a summary in Sect.\ \ref{sect:summary}. Throughout the
paper we assume $(\Omega_m,\Omega_\Lambda)=(0.3,0.7)$ and
$H_0=100h~\mathrm{km~s}^{-1}~\mathrm{Mpc}^{-1}$. 

\section{Data}
\label{sect:data}
We use the COMBO-17 survey for our investigation. COMBO-17 is an acronym for
"Classifying Objects by Medium-Band Observations in 17 filters". This acronym
already contains the most special aspect of COMBO-17: observations in five
broad-band filters ($UBVRI$) and 12 medium-band filters are used to derive
reliable classification and redshift estimates for objects down to
$R=24$. Although in \citet{kleinheinrich2004} it was shown that the
constraints on dark matter halos of galaxies are not significantly improved
compared to a survey with just the five broad-band filters, the 17 filters are
most helpful in obtaining redshifts for the source galaxy population which
are important for translating the measured shear into unbiased mass
estimates. Furthermore, restframe luminosities are derived and used to split the
lens sample into subsamples with intrinsically blue and red colors.

COMBO-17 allows us to study lens galaxies at redshifts around
$z_\mathrm{d}=0.4$ in three disjoint fields of a quarter square degree
each. Only one of the survey fields is a random field. One field is centered
on the Chandra Deep Field South, another one was chosen to study the
supercluster system Abell 901a/b, Abell 902 and represents thus an overdense
line-of-sight. In Sect.\ \ref{sect:clusters} we will investigate the influence
of the supercluster system on the galaxy-galaxy lensing measurement. The
average of these three fields should not be too far from cosmic average. 
 
The data set used here is exactly the same as in the method-focused companion
paper \citet{kleinheinrich2004}. Therefore, we refer the reader to this paper
for more details on the COMBO-17 survey and the shape measurements.

\section{Method}
\label{sect:method}
We use the maximum-likelihood method used by \cite{schneider1997} for
measuring galaxy-galaxy lensing.  Assuming a specific lens model we
can compute the shear $\gamma_{ij}$ on each source $j$ from each lens
$i$. In practise, we use only those lenses whose projected distance
$r$ at the redshift of the lens does not exceed a given
$r_\mathrm{max}$. Sources too close to the field boundaries that might
be lensed by galaxies outside the area of the data set are excluded.

Because the shear from individual galaxies is weak we can sum the shear
contribution from different lenses to derive the total shear acting on source
$j$:
\begin{equation}
\label{eq:method_gamma}
  \gamma_j=\sum_i \gamma_{ij} .
\end{equation}

From this and the observed, psf-corrected ellipticity $\epsilon_j$ the
intrinsic ellipticity $\epsilon_j^{(s)}$ ,
\begin{equation}
\label{eq:method_epsilon}
\epsilon_{j}^{(s)}=\epsilon_{j}-\gamma_{j} ,
\label{eq:epsilon_intr}
\end{equation}
is estimated. The probability for observing this intrinsic ellipticity is
given by
\begin{equation}
\label{eq:method_prob}
P(\epsilon_{j}^{(s)})=\frac{1}{\pi\sigma_{\epsilon}^{2}} \exp{\left[-\frac{|\epsilon_{j}^{(s)}|^{2}}{\sigma_{\epsilon}^{2}}\right]}
\label{eq:P}
\end{equation}
where $\sigma_\epsilon$ is the width of the intrinsic ellipticity
distribution. Multiplying the probabilities from all sources gives the
likelihood for a given set of parameters of the lens model. $\sigma_\epsilon$
is estimated for each source individually depending on its signal-to-noise and
half-light radius, see \citet{kleinheinrich2004} for details. Typical values
are $\sigma_\epsilon\approx 0.4$. 

\subsection{Lens and source selection}
With the COMBO-17 data set we can select lenses and sources based on their
redshifts. We only use objects classified as galaxies or likely galaxies by
COMBO-17. Both lenses and sources lie in the magnitude range $R=18-24$
(Vega). The bright end avoids saturation while the faint end is given by the
magnitude limit down to which the classification and redshift estimation in
COMBO-17 works reliably. Note that due to the redshifts we are able to include
faint lenses and bright sources.

Our lens sample consists of all galaxies with measured redshifts
$z_\mathrm{d}=0.2-0.7$. Lenses with higher redshifts would not add
much to the constraints because only a small number of sources would
lie behind them. Lenses at smaller redshifts are excluded because for
them a given $r_\mathrm{max}$ corresponds to a fairly large angular
separation $\theta$. Because we exclude all sources for which lenses
might lie outside the field boundaries we would therefore have to
exclude too many sources.

Sources lie in the redshift range $z_\mathrm{s}=0.3-1.4$. We only use sources
with individual redshift estimates although in principle we could also include
sources for which only statistical redshifts are available. However, in
\citet{kleinheinrich2004} we found that the inclusion of these sources does
not improve the constraints, probably because sources that are too faint to
get individual redshift estimates also have too noisy shape measurements. A
potential lens-source pair is only evaluated if
$z_\mathrm{s}>z_\mathrm{d}+0.1$. The minimum redshift difference of $dz=0.1$
is chosen to remove pairs where, due to redshift errors, the source is
actually in front of the lens or physically close to it. Thus, we minimize the
contamination from intrinsic alignments \citep{heymans2003,king2002}.

We want to probe lens galaxies out to some projected, physical distance
$r_\mathrm{max}$ from the center. Using the lens redshift, this can be
converted to an angular separation $\theta$ that can vary for different
lens-source pairs. When working with the SIS model (Sect.\ \ref{sect:sis}) we
use $r_\mathrm{max}=150h^{-1}~\mathrm{kpc}$ which we find maximizes the
significance of the measurement. When modelling the lenses by NFW profiles we
extend the maximum separation to $r_\mathrm{max}=400h^{-1}~\mathrm{kpc}$ to
ensure that the region around the virial radius is probed even for the most
massive galaxies.

Shapes can only be measured reliably if objects do not have close
neighbours. Therefore, we only use lens-source pairs with a minimum angular
separation $\theta=8\arcsec$. At $z_\mathrm{d}=0.2$ this translates into a
physical separation $r=18.5h^{-1}~\mathrm{kpc}$, at $z_\mathrm{d}=0.7$ into
$r=40h^{-1}~\mathrm{kpc}$.

In addition to measuring properties averaged over all lenses, we will also
consider blue and red subsamples based on rest-frame colours. The definitions
of 'red' and 'blue' are based on the 'red Sequence' of galaxies in the
color-luminosity plane, found to be present to $z\sim 1$ in COMBO-17. Galaxies
with $<U-V>\leq 1.15-0.31\times z-0.08(M_V-5~log~h +20)$ define the blue
sample while all other galaxies are in the red sample \citep{bell2004}. In
total, our data set contains 11230 blue and 2580 red lens candidates. The
average redshift and SDSS $r$-band rest-frame luminosities are
$\langle z_\mathrm{d} \rangle = 0.46$ and $\langle L \rangle
=0.41\times 10^{10}L_{\sun}$ for the blue sample and $\langle
z_\mathrm{d} \rangle = 0.44$ and $\langle L \rangle =1.27\times
10^{10}L_{\sun}$ for the red sample, respectively.

\section{Results from the SIS model}
\label{sect:sis}
\subsection{Model}
The density distribution of the SIS is given by
\begin{equation}
\label{eq:sis_rho}
   \rho(r)=\frac{\sigma_v^{2}}{2\pi G} \frac{1}{r^2} 
\end{equation}
where $\sigma_v$ is the velocity dispersion of a galaxy. In our
galaxy-galaxy lensing analysis we have to average over many lenses with
different luminosities. However, from the Faber-Jackson and Tully-Fisher
relation it is known that more luminous galaxies have larger velocity
dispersions or rotation velocities, respectively -- at least in the inner few
tens of kpc. Therefore, we assume the following scaling relation between 
velocity dispersion and luminosity:
\begin{equation}
\label{eq:sis_tf}
    \frac{\sigma_v}{\sigma_*}=\left(\frac{L}{L_*}\right)^\eta
\end{equation}
where $\sigma_*$ is the velocity dispersion of an $L_*$ galaxy. We use
$L_*=10^{10}L_{\sun}$ measured in the SDSS $r$-band. For the
Faber-Jackson and Tully-Fisher relations, $\eta\approx 1/4$ or
$\eta\approx 1/3$ is found.

The (aperture) mass-to-light ratio of the galaxies is determined by $\eta$:
\begin{equation}
\label{eq:sis_ml}
    \frac{M(r\leq R)}{L}=\frac{2\sigma_\star^2 (L/L_\star)^{2\eta}R}{GL}\propto L^{2\eta -1}~.
\end{equation}
So for $\eta=0.25$ one has $M/L\propto L^{-0.5}$ inside a fixed aperture while
$M/L\propto L^0=\mathrm{const}$ requires $\eta=0.5$.

For each lens-source pair the shear is given by
\begin{equation}
\label{eq:sis_gamma}
    \gamma(r)=\frac{2\pi\sigma_{v}^{2}}{c^{2}}\frac{D_{\mathrm{ds}}}{D_{\mathrm{s}}}\frac{1}{\theta} .
\end{equation}

\subsection{Results for all galaxies}
The left panel of Fig.\ \ref{fig:sis} shows likelihood contours when all
lenses are used. The best-fit parameters with 1-$\sigma$ error bars are
$\sigma_\star=156^{+18}_{-24}~\mathrm{km/s}$ and $\eta=0.28^{+0.12}_{-0.09}$.
They agree very well with values from the Tully-Fisher or Faber-Jackson
relation, although the scale probed here is an order of magnitude larger. The
(aperture) mass-to-light ratio scales with luminosity as $M/L\propto
L^{-0.44}$. The mass-to-light ratio at fixed radius therefore decreases with
increasing luminosity. 

We also tried values of $r_\mathrm{max}$ larger than
$150h^{-1}~\mathrm{kpc}$. For $r_\mathrm{max}=250h^{-1}~\mathrm{kpc}$
we find $\sigma_\star=138^{+18}_{-18}~\mathrm{km/s}$ and
$\eta=0.31^{+0.12}_{-0.12}$ while for
$r_\mathrm{max}=400h^{-1}~\mathrm{kpc}$ we find
$\sigma_\star=120^{+18}_{-30}~\mathrm{km/s}$ and
$\eta=0.40^{+0.21}_{-0.15}$. There is a clear decrease in
$\sigma_\star$ when larger scales of the galaxy halos are
probed. Further, we see a systematic increase of $\eta$ with
scale. These trends of $\sigma_\star$ and $\eta$ with scale show that
the model adopted here to describe the lenses is too simple. Several
causes are possible: the density profiles of dark matter halos might
decline more steeply than $r^{-2}$, the environment of the lens
galaxies might contribute differently on different scales or the
contribution of different subclasses of lenses might change with
scale. This all clearly shows that when comparing results from
different galaxy-galaxy lensing studies it is important to compare
measurements obtained at similar scales.

\subsection{Results for blue/red subsamples}
\begin{figure*}
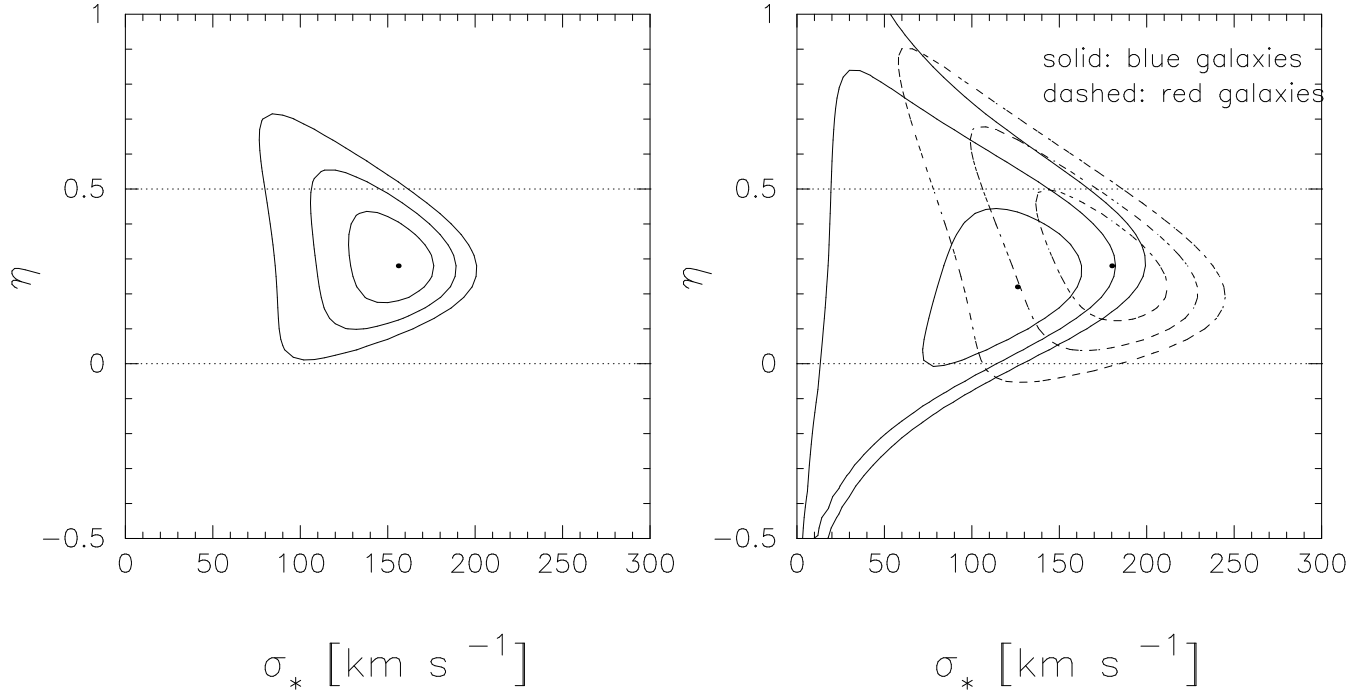

  \begin{center}
   \leavevmode
   \begin{minipage}[l]{0.49\textwidth}  
      \includegraphics[angle=-90,width=\hsize]{figure1.ps}
   \end{minipage}
   \begin{minipage}[r]{0.49\textwidth}
      \includegraphics[angle=-90,width=\hsize]{figure2.ps}
   \end{minipage}
   \caption{Constraints on the velocity dispersion $\sigma_\star$ and the
     Tully-Fisher index $\eta$ for the SIS model using all lenses (left panel)
     and subsamples of blue and red lenses (right panel). Contours are 1-, 2-
     and 3-$\sigma$.}
   \label{fig:sis}
   \end{center}
\end{figure*}

Now we split the lens sample into blue and red subsamples according to the
definition from \citet{bell2004}. Results are shown in the right panel
of Fig.\ \ref{fig:sis} and in Table \ref{tab:sis}. The red sample contains
only 20\% of all lenses but yields even tighter constraints than the 4 times
larger blue sample. This clearly demonstrates that the galaxy-galaxy lensing
signal is dominated by red galaxies. Furthermore, red galaxies have a larger
velocity dispersion at fixed luminosity than blue galaxies. The best-fit
velocity dispersion is about 40\% larger for red galaxies implying that their
(aperture) mass is about twice that of blue galaxies. The scaling $\eta$ does
not differ between the two subsamples. For both red and blue galaxies, the
mass-to-light ratio at fixed radius decreases with increasing luminosity.  

\begin{table*}[htbp]
\begin{center}
\caption{Results for the SIS model for different lens samples. $N_\mathrm{d}$,
  $N_\mathrm{s}$ and $N_\mathrm{p}$ are the numbers of lenses, sources and
  pairs in each measurement. Note that the number of sources can vary for
  different lens selection because not all source candidates are always lying
  within $r_\mathrm{max}$ of a lens. The best fit values of $\sigma_*$ and
  $\eta$ are given with 1-$\sigma$ error bars. The last two columns give the
  aperture mass within $150h^{-1}~\mathrm{kpc}$ and the corresponding
  mass-to-light ratio.}
\begin{tabular}{lrrrrrrr}
\hline\hline
\noalign{\smallskip}
lenses & $N_\mathrm{d}$ & $N_\mathrm{s}$ & $N_\mathrm{p}$ &
$\sigma_\star~[\mathrm{km~s}^{-1}]$ & $\eta$ & $M_\star$
[$10^{12}h^{-1}M_{\sun}$] & $M_\star/L$\\
\noalign{\smallskip}
\hline
\rule[-2mm]{0mm}{0mm}all  & 12167 & 17640 & 105500 & $156^{+18}_{-24}$ &
$0.28^{+0.12}_{-0.09}$ & $1.70^{+0.41}_{-0.48}$ & $170^{+41}_{-48}$\\
\rule[-2mm]{0mm}{0mm}blue &  9875 & 17335 &  83903 & $126^{+30}_{-36}$ & $0.22^{+0.15}_{-0.15}$ & $1.11^{+0.59}_{-0.54}$ & $111^{+59}_{-54}$\\
\rule[-2mm]{0mm}{0mm}red  &  2292 & 11063 &  21597 & $180^{+24}_{-30}$ & $0.28^{+0.15}_{-0.12}$ & $2.26^{+0.64}_{-0.69}$ & $226^{+64}_{-69}$\\
\hline
\end{tabular}
\label{tab:sis}
\end{center}
\end{table*}

\section{Results from the NFW profile}
\label{sect:NFW}
\subsection{Model}
\label{sect:nfw_model}
The density distribution of the NFW profile is given by
\begin{equation}
\label{eq:nfw_rho}
  \rho(r)=\frac{\delta_{c}~\rho_{c}}{(r/r_{s})(1+r/r_{s})^{2}}~,
\end{equation}
where $\delta_c$ is a characteristic density, $\rho_{c}$ is the critical 
density for a closed universe and $r_{s}$ is a scale radius. At $r\sim r_s$ 
the density profile turns from $\rho(r)\propto r^{-1}$ to $\rho(r)\propto
r^{-3}$. 

The virial radius $r_\mathrm{vir}$ is defined as the radius inside which the
mean density is 200 times the mean density of the universe. The mass inside
the virial radius is the virial mass
\begin{equation}
\label{eq:nfw_M_vir}
   M_\mathrm{vir}=\frac{800}{3}\pi \rho_m r_{vir}^{3}
\end{equation}
with $\rho_m=\Omega_m \rho_c$.
The ratio between virial radius und scale radius is the concentration 
\begin{equation}
\label{eq:nfw_c}
   c=r_\mathrm{vir}/r_s ~.
\end{equation}
From Eqs.\ (\ref{eq:nfw_rho})-(\ref{eq:nfw_c}) follows the relation between
the characteristic density and the concentration
\begin{equation}
\label{eq:nfw_delta_c}
   \delta_{c}=\frac{200~\Omega_m}{3}~\frac{c^{3}}{\ln(1+c)-c/(1+c)}.
\end{equation}

In analogy to the Tully-Fisher and Faber-Jackson relations, which imply a
relation between luminosity and mass, we impose a similar relation between the
virial radius (a proxy for the virial mass) and luminosity here,
\begin{equation}
\label{eq:nfw_tf}
  \frac{r_\mathrm{vir}}{r_\mathrm{vir}^*}=\left(\frac{L}{L_*}\right)^\alpha
\end{equation}

The virial mass-to-light ratio is thus given by
\begin{equation}
\label{eq:nfw_ml}
    \frac{M_\mathrm{vir}}{L}=\frac{800}{3}\pi \rho_m \frac{r_{vir}^{3}}{L}\propto L^{3\alpha -1}~.
\end{equation}
A constant mass-to-light ratio $M/L\propto L^0=\mathrm{const}$ requires $\alpha=1/3$.

The shear from the NFW model is calculated in
\citet{bartelmann1996} and \citet{wright2000}. 

\subsection{Constraints on the concentration $c$}
\begin{figure}
  \begin{center}
   \leavevmode
      \includegraphics[angle=-90,width=\hsize]{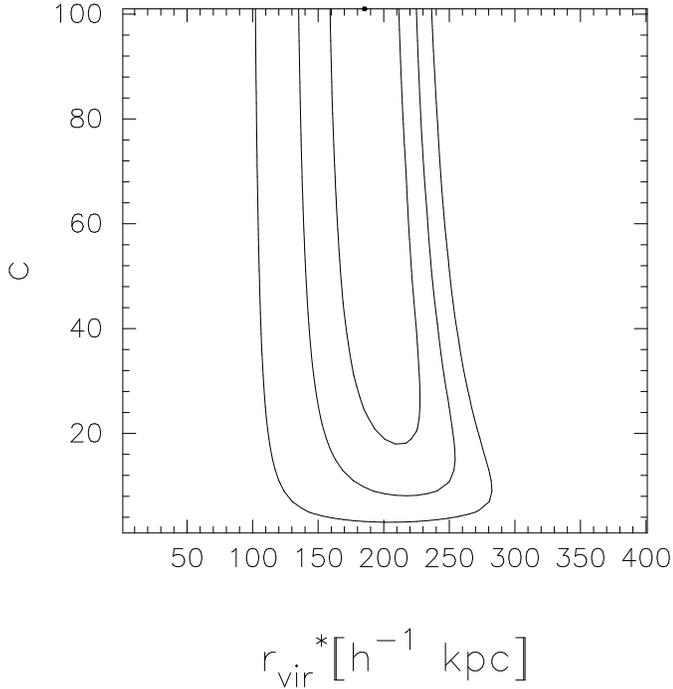}
   \caption{Constraints on the virial radius $r_\mathrm{vir}^*$ and the
      concentration $c$ for the NFW profile obtained from the full lens
      sample with $\alpha=0.3$ fixed.}
   \label{fig:nfw_c}
   \end{center}
\end{figure}

First, we try to constrain the virial radius $r_\mathrm{vir}^*$ and the
concentration $c$ for a fixed $\alpha$. Figure \ref{fig:nfw_c} shows results
for the whole lens sample with $\alpha=0.3$. Using $\alpha=0$ instead does
not change the shape of the contours but just shifts them toward lower values
of $r_\mathrm{vir}^*$. The dependence on $c$ is weak and only a lower
bound can be obtained. The 1-$\sigma$ lower limit for $c$ alone is $c>29$, the
2-$\sigma$ lower limit is $c>11$. In the following we will use a fixed
$c=20$. This choice is somewhat arbitrary. However, smaller values are
disfavoured by our data and would increase the error bars of the following
measurements. Larger values, on the other hand, would hardly change the
results. 

We will later find virial radii of about $r_\mathrm{vir}^*\approx
200h^{-1}~\mathrm{kpc}$. With $c=20$ this gives a scale radius
$r_s\approx 10h^{-1}~\mathrm{kpc}$. Also for smaller concentrations
$c=10$ or $c=5$, $r_s$ would only increase to about
$20h^{-1}~\mathrm{kpc}$ or $40h^{-1}~\mathrm{kpc}$
respectively. However, as detailed in Sect.\ \ref{sect:method},
$(20-40)h^{-1}~\mathrm{kpc}$ is just the minimum distance from the
center where we start to probe the lens galaxies. Therefore, we cannot
expect to be sensitive to the scale radius $r_s$ and thus to the
concentration $c$.

Note that the value of $c$ is dependent on our definition of the
virial radius. Unfortunately, there is no unique definition of the
virial radius in the literature. Sometimes it is referred to as the
radius inside which the mean density is some over-density times the
{\em mean} density of the universe, and sometimes as the radius inside
which the mean density is some over-density times the {\em critical}
density of the universe. Changing the definition of the virial radius
would change relation (\ref{eq:nfw_delta_c}) and therefore $c$. The
shape of the density profile remains unaffected so that the values of
the scale radius $r_s$ and the characteristic density $\delta_c$ do
not change. Defining the virial radius as the radius inside which the
mean density is 200 times the critical density instead of our
definition from Sect.\ \ref{sect:nfw_model} would lower the
concentration from $c=20$ to approximately $c'=12.4$. The virial
radius would therefore become about
$r_\mathrm{vir}'=0.62r_\mathrm{vir}$ and the virial mass
$M_\mathrm{vir}'=0.79M_\mathrm{vir}$. We emphasize these differences
here only for clarity and easier comparability to other works. All
figures and results presented in this paper use the definition of the
virial radius given in Sect.\ \ref{sect:nfw_model}.

\subsection{Results for all galaxies}
Using all lenses we obtain the likelihood contours shown in the left panel of
Fig.\ \ref{fig:nfw}. The best-fit parameters with 1-$\sigma$ error bars are
$r_\mathrm{vir}^*=209^{+24}_{-32}h^{-1}~\mathrm{kpc}$ and
$\alpha=0.34^{+0.12}_{-0.16}$. The corresponding virial mass is
$M_\mathrm{vir}^*=6.4^{+2.5}_{-3.3}\times 10^{11}h^{-1}M_{\sun}$. This mass
estimate is considerably smaller than the mass estimate from the SIS model
($M_\mathrm{SIS}(r\leq 150h^{-1}~\mathrm{kpc})=1.70^{+0.41}_{-0.48}\times
10^{12}h^{-1}M_{\sun}$, see Table \ref{tab:sis}), although the virial mass
even encloses a larger radius than the aperture used for the SIS
model. \citet{wright2000} already pointed out that for galaxy-sized halos the
SIS model yields much larger mass estimates than the NFW model. Decreasing
$r_\mathrm{max}$ from $r_\mathrm{max}=400h^{-1}~\mathrm{kpc}$ to
$250h^{-1}~\mathrm{kpc}$ or $150h^{-1}~\mathrm{kpc}$ yields larger
$r_\mathrm{vir}^*$ and smaller $\alpha$,
$r_\mathrm{vir}^*=233^{+24}_{-32}h^{-1}~\mathrm{kpc}$ and
$\alpha=0.26^{+0.12}_{-0.08}$ for
$r_\mathrm{max}=150h^{-1}~\mathrm{kpc}$. Unlike the SIS model, the
differences here are within the 1-$\sigma$ uncertainties. Therefore, it seems
that the NFW profile provides a better fit to the data than the SIS
model. However, the tendency to larger virial radii for decreasing
$r_\mathrm{max}$ might indicate that our modelling of NFW profiles 
also needs refinements. 

For the NFW model with $r_\mathrm{max}=400h^{-1}~\mathrm{kpc}$ we obtain a
scaling of the virial mass-to-light ratio with luminosity as $M/L\propto
L^{0.02}$ implying almost the same mass-to-light ratio for all
luminosities. In contrast, we find a decreasing mass-to-light ratio at
fixed radius with increasing luminosity for the SIS model. However, the
scaling relation found from the SIS model is only marginally excluded at the
1-$\sigma$ level by the measurement from the NFW  model.

\subsection{Results for blue/red subsamples}
\begin{figure*}
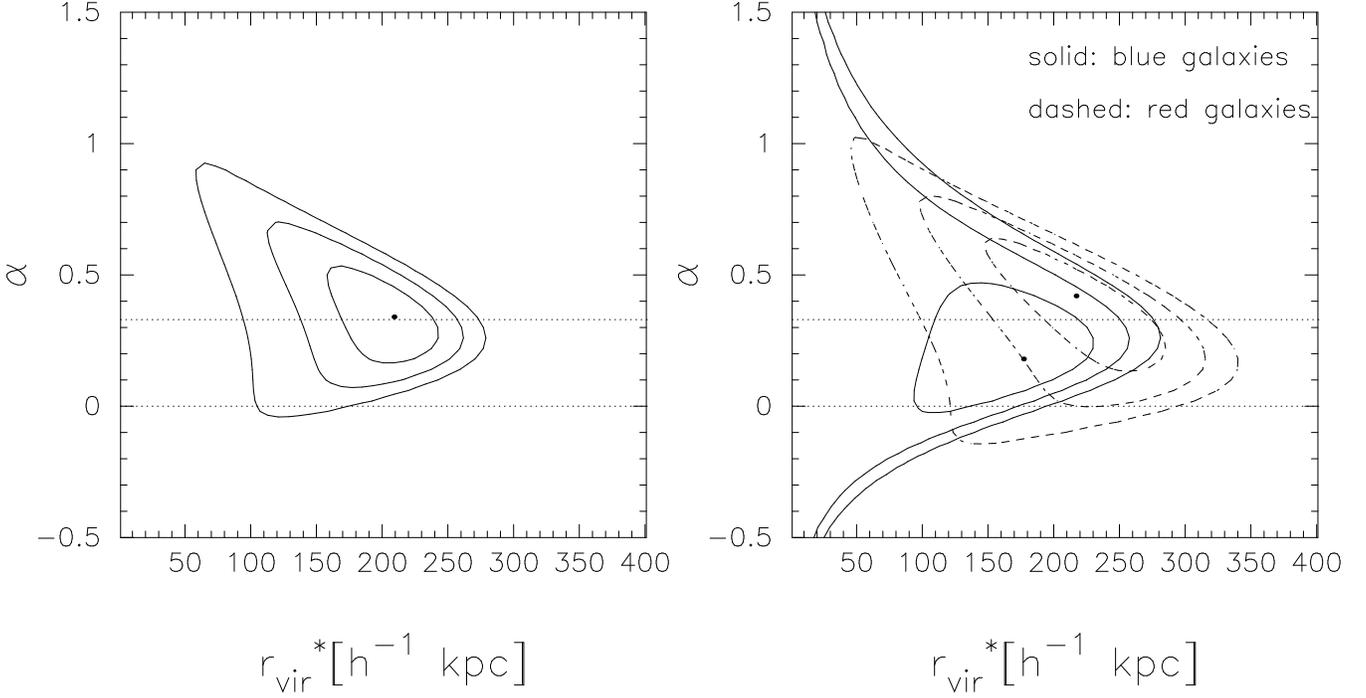

  \begin{center}
   \leavevmode
   \begin{minipage}[l]{0.49\textwidth}  
      \includegraphics[angle=-90,width=\hsize]{figure4.ps}
   \end{minipage}
   \begin{minipage}[r]{0.49\textwidth}
      \includegraphics[angle=-90,width=\hsize]{figure5.ps}
   \end{minipage}
   \caption{Constraints on the virial radius $r_\mathrm{vir}^*$ and its scaling
     with luminosity ($\alpha$) for the NFW profile using all lenses (left
     panel) and subsamples of blue and red lenses (right panel). Contours are
     1-, 2- and 3-$\sigma$.}
   \label{fig:nfw}
   \end{center}
\end{figure*}

Again, we split the lens sample into a blue and a red subsample; see the
right panel of Fig.\ \ref{fig:nfw} and Table \ref{tab:nfw}. We find that red
galaxies have a larger virial radius at a given luminosity $L_\star$ and a
larger $\alpha$ than blue galaxies. However, the significance of these
differences is only about 1 $\sigma$. The larger virial radius for the red
galaxies implies a virial mass that is almost a factor of 2 larger than that
of blue galaxies. Even within the virial radius of the blue galaxies, the mass
of red galaxies is much larger,
$M_\mathrm{red}(r\leq177h^{-1}~\mathrm{kpc})=6.5^{+5.1}_{-3.1}\times10^{11}h^{-1}M_{\sun}$.
Formally, the best-fit mass-to-light ratios scale with luminosity as
$M/L\propto L^{-0.46}$ for blue galaxies and as $M/L\propto L^{0.26}$ for red
ones, but for both subsamples a mass-to-light ratio independent of luminosity
is consistent with the data.

\begin{table*}[htbp]
\begin{center}
\caption{Constraints on dark matter halos of galaxies modelled by NFW profiles
  for different lens samples. $N_\mathrm{d}$, $N_\mathrm{s}$ and
  $N_\mathrm{p}$ are the numbers of lenses, sources and pairs in each
  measurement. These numbers differ from those given in Table \ref{tab:sis}
  because of the larger $r_\mathrm{max}$ used here. The virial radius
  $r_\mathrm{vir}^*$ and $\alpha$ are fitted quantities (see Fig.\
  \ref{fig:nfw}), the virial mass $M_\mathrm{vir}^*$ and the virial
  mass-to-light ratio $M_\mathrm{vir}^*/L$ are calculated from
      $r_\mathrm{vir}^*$. $\beta=3\alpha-1$ gives the scaling of
      $M_\mathrm{vir}^*/L$ with luminosity, $M_\mathrm{vir}^*/L\propto
      L^\beta$. All errors are 1-$\sigma$.}
\begin{tabular}{lrrrccccr}
\hline\hline
\noalign{\smallskip}
lenses & $N_\mathrm{d}$ & $N_\mathrm{s}$ & $N_\mathrm{p}$  &
$r_\mathrm{vir}^*$ & $\alpha$ & $M_\mathrm{vir}^*$ & $M_\mathrm{vir}^*/L$ &
$\beta$\\ 
 & & & & [$h^{-1}~\mathrm{kpc}$] & & [$10^{11}h^{-1}M_{\sun}$] &
 [$h(M/L)_{\sun}$] &\\ 
\noalign{\smallskip}
\hline
\rule[-2mm]{0mm}{0mm} all & 11311 & 13956 & 566466 & $209^{+24}_{-32}$ &
$0.34^{+0.12}_{-0.16}$ & $6.4^{+2.5}_{-3.3}$ & $64^{+25}_{-33}$ &
$0.02^{+0.36}_{-0.48}$\\ 
\rule[-2mm]{0mm}{0mm} blue & 9181 & 13936 & 451420 & $177^{+40}_{-48}$ &
$0.18^{+0.20}_{-0.12}$ & $3.9^{+3.3}_{-2.4}$ & $39^{+33}_{-24}$ &
$-0.46^{+0.60}_{-0.36}$\\ 
\rule[-2mm]{0mm}{0mm} red &  2130 & 13603 & 115046 & $217^{+56}_{-48}$ &
$0.42^{+0.16}_{-0.20}$ & $7.1^{+7.1}_{-3.8}$ & $71^{+71}_{-38}$ &
$0.26^{+0.48}_{-0.60}$\\\hline 
\hline
\end{tabular}
\label{tab:nfw}
\end{center}
\end{table*}

\section{Influence of clusters}
\label{sect:clusters}

One of the COMBO-17 fields -- the A 901 field -- has been chosen specifically
to study the supercluster composed of the components Abell 901a, 901b and 902
at a redshift of $z=0.16$ \citep{gray2002,gray2004}. Later, another cluster
(named CBI in \citet{taylor2004}) was detected behind Abell 902 using the 3-D
distribution of galaxies. The masses of these four clusters were measured
jointly using weak lensing \citep{taylor2004}. Table \ref{tab:cluster_data}
shows central positions, redshifts and velocity dispersions for the different
clusters modelled as SISs. We address in this section the influence of these
clusters on our galaxy-galaxy lensing measurements. Two measurements are
compared: one ignoring the presence of the clusters and one including the
shear from the clusters $\gamma_\mathrm{cl}$. In the second case, Eq.\
(\ref{eq:method_epsilon}) becomes
\begin{equation}
\epsilon_{j}^{(s)}=\epsilon_{j}-\gamma_{j}-\gamma_\mathrm{cl}~.
\label{eq:cluster_epsilon}
\end{equation}
The shear of the clusters is computed assuming isothermals spheres (see Eq.\
(\ref{eq:sis_gamma})) with the parameters from Table
\ref{tab:cluster_data}. The lens galaxies are also modelled as isothermal
spheres, see Sect.\ \ref{sect:sis}. 
Instead of using the shear $\gamma$ of the clusters we also tried using the
reduced shear $g=\gamma/(1-\kappa)$ and including the magnification
$\mu=((1-\kappa)^2-\gamma^2)^{-1}$ to correct the luminosities of the
lenses, the convergence $\kappa$ here, being calculated from the SIS fits of the
clusters. We found that the difference to just using $\gamma$ in both 
cases is negligible.

The S 11 field contains the cluster Abell 1364 at a redshift $z=0.11$
for which we fit a velocity dispersion
$\sigma=615^{+110}_{-140}~\mathrm{km/s}$. As for the foreground
clusters in the A 901 field, we compare measurements for the S 11
field, first including and then ignoring the cluster shear. We find
that the cluster shear is negligible for the S 11 field.

\begin{table}[htbp]
\begin{center}
\caption{Central positions, redshifts and velocity dispersions of the known
  clusters in the A 901 field \citep{taylor2004}.}  
\begin{tabular}{l|c|c|c|c}
\hline\hline
\noalign{\smallskip}
 & $\alpha_{\mathrm{J2000}}$ & $\delta_{\mathrm{J2000}}$ & $z$ & $\sigma_v$ [km/s]\\
\noalign{\smallskip}
\hline
\rule[-2mm]{0mm}{0mm}A901a & $09^{\mathrm{h}}56^{\mathrm{m}}26.4^{\mathrm{s}}$
& $-09^{\mathrm{h}}57^{\mathrm{m}}21.7^{\mathrm{s}}$ & $0.16$ & $680^{+25}_{-90}$ \\
\rule[-2mm]{0mm}{0mm}A901b & $09^{\mathrm{h}}55^{\mathrm{m}}57.4^{\mathrm{s}}$
& $-09^{\mathrm{h}}59^{\mathrm{m}}02.7^{\mathrm{s}}$ & $0.16$ & $600^{+40}_{-85}$\\
\rule[-2mm]{0mm}{0mm}A902 & $09^{\mathrm{h}}56^{\mathrm{m}}33.6^{\mathrm{s}}$
& $-10^{\mathrm{h}}09^{\mathrm{m}}13.1^{\mathrm{s}}$ & $0.16$ & $470^{+100}_{-280}$\\
\rule[-2mm]{0mm}{0mm}CBI & $09^{\mathrm{h}}56^{\mathrm{m}}39.6^{\mathrm{s}}$
& $-10^{\mathrm{h}}10^{\mathrm{m}}21.6^{\mathrm{s}}$ & $0.48$ & $730^{+160}_{-340}$\\
\end{tabular}
\label{tab:cluster_data}
\end{center}
\end{table}

\subsection{Influence of foreground clusters at $z=0.16$}
\begin{figure*}[htbp]
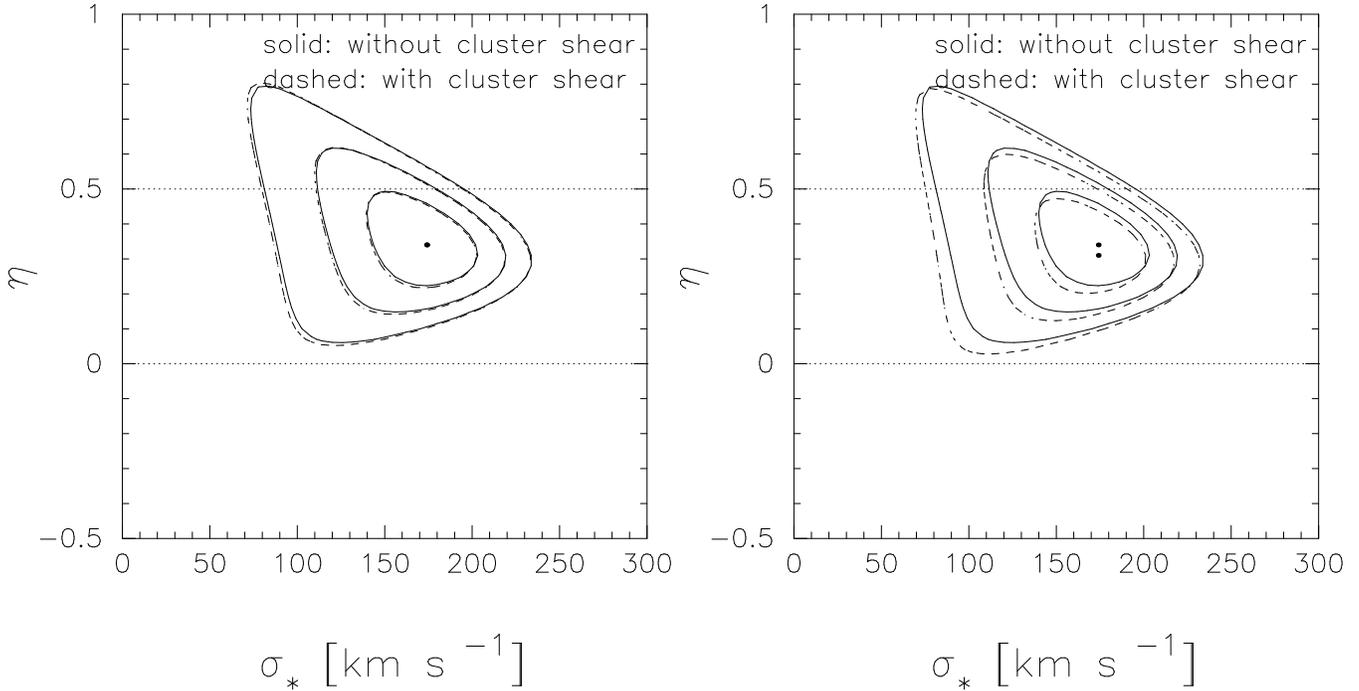

  \begin{center}
   \leavevmode
   \begin{minipage}[l]{0.49\textwidth}  
      \includegraphics[angle=-90,width=\hsize]{figure6.ps}
   \end{minipage}
   \begin{minipage}[r]{0.49\textwidth}
      \includegraphics[angle=-90,width=\hsize]{figure7.ps}
   \end{minipage}
   \caption{Influence of the clusters shear on the galaxy-galaxy lensing
     measurement. Solid lines refer to measurements ignoring the cluster
     shear, dashed lines to measurements including the cluster shear. In the
     left panel, only the foreground clusters Abell 901a/b and Abell 902 are
     taken into account. In the right panel also the background cluster CBI is
     used.}
   \label{fig:clusters}
   \end{center}
\end{figure*} 

Only galaxies with redshifts $0.2<z_\mathrm{d}<0.7$ have been selected as lens
galaxies. Therefore, the lens sample should contain no galaxies that lie in
the foreground clusters (Abell 901a/b, Abell 902). The expectation is that the
foreground clusters will not influence the galaxy-galaxy lensing measurements
because the shear from the foreground clusters will be in random directions
with respect to the orientations of the lens-source pairs used for
galaxy-galaxy lensing.

The left panel of Fig.\ \ref{fig:clusters} shows likelihood contours obtained
from the A 901 field both ignoring and including the shear from the
foreground clusters Abell 901a/b and Abell 902. The shear from CBI is ignored.
Hardly any difference is seen between both cases, from which we measure 
$\sigma_\star=174^{+24}_{-24}~\mathrm{km/s}$ and $\eta=0.34^{+0.12}_{-0.09}$
(1-$\sigma$ errors). 

\begin{figure}[htbp]
  \begin{center}
   \leavevmode
      \includegraphics[width=\hsize]{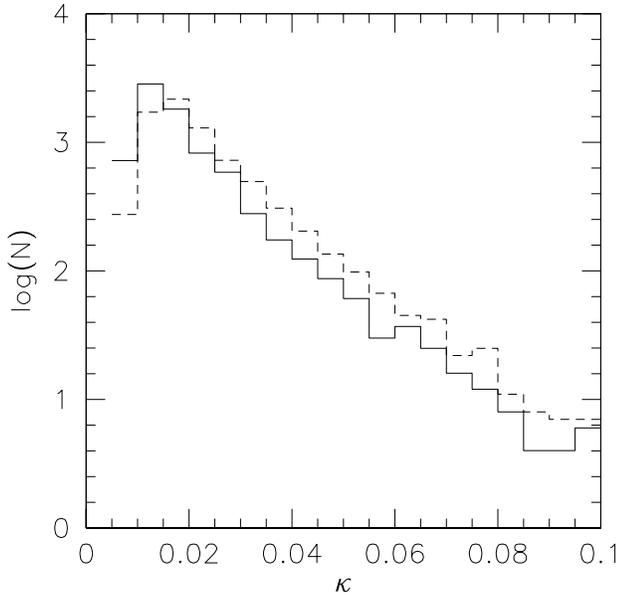}
   \caption{Histogram over $\kappa$ due to the foreground clusters only
      (solid) and due to the foreground clusters and CBI together (dashed)}
   \label{fig:clusters_kappa}
   \end{center}
\end{figure} 

As a second test we ignore the cluster shear but exclude all sources
from the galaxy-galaxy lensing measurement for which the convergence
$\kappa$ (and thus the shear $\gamma$) from the foreground clusters
exceeds some threshold. Figure \ref{fig:clusters_kappa} shows that
from the foreground clusters alone $\kappa<0.1$ for almost all
sources. The maximum is $\kappa=0.17$. Therefore the weak shear limit
is still valid in Eq.\ (\ref{eq:cluster_epsilon}). When excluding all
sources with $\kappa>0.05$ (224 out of 6481 sources), the difference
to the measurement using all sources is negligible. When excluding
further sources, the contours start to widen significantly but
maintain the same minimum. This widening can be explained by the
decreasing number statistics. In particular, we do not see any hint of
a bias from the clusters. When excluding the 22\% of sources with
$\kappa>0.025$, the 3-$\sigma$ contour remains closed. The best-fit
parameters with 1-$\sigma$ errors are then
$\sigma_\star=180^{+24}_{-30}~\mathrm{km/s}$ and $\eta=0.31\pm0.12$.

\subsection{Influence of CBI at $z=0.48$}
The cluster CBI has a redshift $z=0.48$ and therefore some of the lens
galaxies will reside in the cluster halo. This implies that additional mass is
present around these galaxies. Ignoring the cluster CBI, this additional mass
will be assigned to the galaxies and thus the derived masses (or velocity
dispersions) will be too high. \citet{guzik2002} modelled this case for the
SDSS data. They find from theoretical models that clusters increase the 
shear measurements of galaxies. The strength of this effect depends on the
distance from the lens center, it increases from zero toward a maximum around
$r=200h^{-1}~\mathrm{kpc}$ and then decreases again. The peak contribution and
the shape of the decrease depend on the details of the model.

The right panel of Fig.\ \ref{fig:clusters} shows the same as the left
panel but this time includes the shear from CBI in the cluster
measurement. Although the change in the contours is clearly visible,
the shear from the background cluster CBI does not seem to have a big
influence; the contours only widen marginally. The best-fit parameters
with their 1-$\sigma$ errors become
$\sigma_\star=174^{+24}_{-24}~\mathrm{km/s}$ and
$\eta=0.31^{+0.12}_{-0.09}$.  Note, however, that the resolution of
our grid in parameter space is only $\Delta \eta=0.03$.
 
As a second test we again ignore the cluster shear and exclude objects with
large $\kappa$ from the foreground clusters and CBI together. Figure
\ref{fig:clusters_kappa} shows a histogram of $\kappa$, the maximum 
of which is $\kappa=0.26$. When excluding all
sources with $\kappa>0.05$ we obtain likelihood contours similar to the case
when including the cluster shear. 345 out of 6481 sources are then
excluded. Excluding further sources results in wider contours. The
3-$\sigma$ contour stays closed when we exclude sources with
$\kappa>0.03$. These are 22\% of all sources and the contours do not
shift compared to the case where the cluster shear is ignored. 

A third test is to exclude all lenses that are close to the center of the
background cluster CBI. We exclude every lens lying within
$1h^{-1}~\mathrm{Mpc}$ projected distance from the cluster center
(corresponding to $\theta=240\arcsec$ angular separation) and with a
redshift difference less than 0.2. These are 221 out of 4444 lenses. The
likelihood contours hardly change when these lenses are excluded.

From all these tests we conclude that our results are not biased from the
presence of several clusters in the A 901 field. This conclusion also holds
when the SIS model is fitted within $r_\mathrm{max}=400h^{-1}~\mathrm{kpc}$
instead of $150h^{-1}~\mathrm{kpc}$. The little influence even the
background cluster CBI has might be surprising but it is understandable given
the small number of lenses that actually reside within this cluster. In
particular we note that the comparably large value of $\sigma_\star$ obtained
from the A 901 field which is about 1-$\sigma$ higher than from all three
fields together is not caused directly by the clusters. Instead, it
seems that cosmic variance is still an issue even on the scales of our
survey fields.

\section{Comparison to other surveys}
\label{sect:comparison}
In this section, we compare our results to results from previous 
galaxy-galaxy lensing studies. Such investigation is hampered by the very
different data sets and techniques used in the different investigations. For
our comparison, we concentrate on results from the RCS \citep{hoekstra2004}
and the SDSS \citep{guzik2002, seljak2004} data.

\subsection{Comparison to the RCS}
\label{sect:comparison_rcs}
The galaxy-galaxy lensing analysis from the RCS \citep{hoekstra2004} is in
many regards similar to the one presented here and thus very suitable for a
comparison of results. Both surveys reach a comparable depth so that one can
expect that comparable sets of lens galaxies are probed. Further,
\citet{hoekstra2004} use the maximum-likelihood technique by
\citet{schneider1997}, model lenses as SISs and by NFW profiles and fit to a
fiducial lens galaxy with a characteristic luminosity $L_\star$, just as we
do. However, some differences exist. The area of the RCS used for the
galaxy-galaxy lensing analysis is $45.5~\mathrm{deg} ^2$, so about 60 times
larger than that used here. On the other hand, \citet{hoekstra2004} only have
observations in a single filter available. Therefore, they have to select
lenses and sources based on magnitude cuts and they have to use redshift
probability distributions for estimating luminosities and for shear
calibration. Furthermore, \citet{hoekstra2004} can only investigate lens galaxies
over an angular scale while we are able to probe the same physical scale of
all lenses. \citet{hoekstra2004} fit their models within $2'$ which they
estimate corresponds to about $350 h^{-1}~\mathrm{kpc}$ at the mean redshift
of the  lenses. This is considerably larger than the region we probe for the
SIS model ($r_\mathrm{max}=150 h^{-1}~\mathrm{kpc}$) and comparable to the
region probed for the NFW profile ($r_\mathrm{max}=400
h^{-1}~\mathrm{kpc}$). Another difference lies in the definition of the
fiducial luminosity. We use $L_\star=10^{10}h^{-2}L_{r,{\sun}}$ measured in the
SDSS $r$-band while \citet{hoekstra2004} use $L_B=10^{10}h^{-2}L_{B,{\sun}}$ as
reference. From our data set we calculate that lenses with
$L_B=10^{10}h^{-2}L_{B,{\sun}}$ have SDSS-$r$-band luminosities of about
$L_r=1.1\times 10^{10}h^{-2}L_{r,{\sun}}$. In \citet{kleinheinrich2004} it is
shown for the SIS that it is not possible to constrain the scaling relation
between velocity dispersion and luminosity without multi-color
data. Therefore, \citet{hoekstra2004} have to assume this scaling relation
unlike being able to fit it as we do.

\subsubsection{SIS model}
For better comparability with the RCS results we redo our fit to the SIS model
with $L_\star=1.1\times 10^{10}h^{-2}L_{r,{\sun}}$ and $r_\mathrm{max}=350
h^{-1}~\mathrm{kpc}$. We obtain
$\sigma_\star=132^{+18}_{-24}~\mathrm{km~s}^{-1}$ and
$\eta=0.37^{+0.15}_{-0.15}$ from the whole lens sample.

\citet{hoekstra2004} do not specifically fit the SIS model using the
maximum-likelihood technique of \citet{schneider1997}. Instead, they fit the
SIS model to the measured galaxy-mass cross-correlation, and they constrain
the truncated SIS model using the maximum-likelihood technique. The velocity
dispersion obtained from the truncated SIS model becomes that of the SIS in
the limit of infinite truncation parameter $s$. In both cases,
\citet{hoekstra2004} assume a scaling relation as in Eq.\ (\ref{eq:sis_tf})
with $\eta=0.3$ and a characteristic luminosity $L_\star$ as detailed in
Sect.\ \ref{sect:comparison_rcs}. From the galaxy-mass cross-correlation
function they obtain $\sigma=140\pm4\pm3~\mathrm{km~s}^{-1}$. The truncated
SIS yields $\sigma=136\pm5\pm3~\mathrm{km~s}^{-1}$. Both results are in good
agreement with our results. Conversely, we are able to confirm the
value of $\eta$ adopted by \citet{hoekstra2004}.

The error bars of \citet{hoekstra2004} are about 5 times smaller than ours. 
From the difference in area between the RCS and COMBO-17 one would even expect
a factor of about $\sqrt{60}\approx 8$ difference. That their error bars are
not that much smaller can be attributed to the detailed classification and
redshifts available in COMBO-17. In \citet{kleinheinrich2004} it is shown
that the error bar on $\sigma_\star$ increases by about 30\% when redshifts are
omitted. The influence of accurate redshift estimates on the determination of
$\eta$ is found to be much more severe -- from just a single passband no
meaningful constraints on $\eta$ can be derived. This explains why we are able
to fit $\eta$ while \citet{hoekstra2004} had to assume a fixed value.

\subsubsection{NFW profile}
For better comparability with the RCS we also redo our fit to the NFW profile
with $L_\star=1.1\times 10^{10}h^{-2}L_{r,{\sun}}$ and $r_\mathrm{max}=350
h^{-1}~\mathrm{kpc}$. We obtain
$r_\mathrm{vir}^*=217^{+24}_{-32}h^{-1}~\mathrm{kpc}$ and 
$\alpha=0.30^{+0.16}_{-0.12}$ from the whole lens sample. This corresponds to
a virial mass $M_\mathrm{vir}^*=7.1^{+2.6}_{-2.7}\times 10^{11}h^{-1}M_{\sun}$.

The exact modelling of \citet{hoekstra2004} is somewhat different from
ours. While we fit the virial radius and its scaling with luminosity,
\citet{hoekstra2004} assume scaling relations between the maximum rotation
velocity and virial mass, respectively, with luminosity and fit the virial
velocity $v_\mathrm{vir}$ and the scale radius $r_s$. The virial
velocity is directly related to the virial mass and virial radius by
$v_\mathrm{vir}=\sqrt{GM_\mathrm{vir}/r_\mathrm{vir}}\propto
r_\mathrm{vir}$. All three quantities are independent of the scale
radius $r_s$. Therefore, we can readily compare our results on the virial
mass to those from \citet{hoekstra2004}. They find
$M_\mathrm{vir}=(8.4\pm 0.7 \pm 0.4)\times 10^{11}h^{-1}M_{\sun}$ in
good agreement with our result. Furthermore, their constraint on the scale
radius $r_s=16.2^{+3.6}_{-2.9}h^{-1}~\mathrm{kpc}$ is roughly
consistent with the scale radius implied by the virial radius we find
and the concentration we assume.

\subsection{Comparison to the SDSS}
\label{sect:comparison_sdss}
While COMBO-17 and the RCS are both deep surveys from which lens galaxies at
redshifts $z_\mathrm{d}\approx 0.4$ can be probed, the SDSS is a much
shallower survey with a correspondingly small average lens redshift. 
To date, several observational weak lensing analyses have been
published investigating different aspects of dark matter halos of galaxies and
using consecutively larger parts of the survey
\citep{fischer2000,mckay2001,seljak2002,guzik2002,sheldon2004,seljak2004}. For
the comparison to our results we will concentrate on the results by
\citet{guzik2002}. 

\citet{guzik2002} use the halo model for their galaxy-galaxy lensing
analysis. This model takes not only the contribution from galaxies
themselves into account but also from the surrounding group and
cluster halos. This of course complicates the comparison. On the other
hand, \citet{guzik2002} find that on average only about 20\% of the
galaxies are non-central galaxies, for which the group/cluster
contribution is important. Therefore, the group/cluster contribution
to our mass estimates is probably well within our error bars.
\citet{guzik2002} only fit NFW profiles to their data. They define the virial
radius as radius inside which the mean density is 200 times the critical
density of the universe. Therefore, we have to multiply our virial masses by a
factor 0.79 to compare them to theirs. Similar to our approach,
\citet{guzik2002} assume a relation between mass and luminosity
$M/M_\star=(L/L_\star)^\beta$. $M_\star$ and $\beta$ are derived from
different passbands. Here, we just compare to the results from the
$r'$-band in which the reference luminosity is $L_\star=1.51\times
10^{10}h^{-2}L_{\sun}$. 

\subsubsection{All galaxies}
Assuming that group/cluster halos only contribute to the faintest lens
galaxies, \citet{guzik2002} find a virial mass $M_\star=(8.96\pm1.59)\times
10^{11}h^{-1}M_{\sun}$ and $\beta=1.51\pm0.16$. Our measurement of
$r_\mathrm{vir}^*$ implies for a galaxy with $L=1.5\times
10^{10}h^{-2}L_{r,{\sun}}$ a virial radius of
$r_\mathrm{vir}=240^{+27}_{-37}h^{-1}~\mathrm{kpc}$. Here, we have used the
measured $\alpha=0.34$ and ignored its errors and still use our definition of
the virial radius. Changing to the definition adopted by \citet{guzik2002}
this corresponds to a virial mass of
$M_\mathrm{vir}'=7.6^{+2.9}_{-3.0}\times10^{11}h^{-1}M_{\sun}$ in very good
agreement with the result from the SDSS. Our measurement of
$\alpha=0.34^{+0.12}_{-0.16}$ corresponds to
$\beta=1.02^{+0.36}_{-0.48}$. This is smaller than the value found by
\citet{guzik2002} but the difference is not very significant. When
\citet{guzik2002} assume the same group/cluster contribution for all
luminosity bins, their best-fit $\beta$ drops to
$\beta=1.34\pm0.17$. Therefore it might well be that the differences between
COMBO-17 and the SDSS concerning the scaling between mass and light is due to
the differences in the modelling.

\subsubsection{Early-/late-type subsamples}
\citet{guzik2002} also split the lens sample into early- and late-type
subsamples. Both samples contain about equal numbers. This already indicates
that these subsamples cannot be too similar to our subsamples of red and blue
galaxies in which only about 20\% of all galaxies belong to the red
subsample. However, we should at least be able to see similar trends from
late-type to early-type galaxies as from blue to red ones. \citet{guzik2002}
fix the scaling of the mass with luminosity to the value derived for the whole
lens sample ($\beta=1.51$) and also adopt the same $L_\star=1.51\times
10^{10}h^{-2}L_{\sun}$ for both subsamples. For late-type galaxies they find a
virial mass of $L_\star$-galaxies of $M_\star=(3.26\pm2.08)\times
10^{11}h^{-1}M_{\sun}$, for early-type galaxies they find
$M_\star=(10.73\pm2.53)\times 10^{11}h^{-1}M_{\sun}$. We scale the virial radii
of blue and red galaxies measured at $L=10^{10}h^{-2}L_{\sun}$ to the higher
$L_\star$ used by \citet{guzik2002} using the measured $\alpha$ for the two
samples and calculate from that the virial masses assuming the same definition
as \citet{guzik2002}. We obtain
$M_\mathrm{vir}'=3.8^{+3.2}_{-2.3}\times10^{11}h^{-1}M_{\sun}$ for the blue
sample and $M_\mathrm{vir}'=9.4^{+9.4}_{-5.0}\times10^{11}h^{-1}M_{\sun}$ for
the red sample. The agreement with the SDSS results is surprisingly good given
the different definitions of the subsamples and the different lens
redshifts. Most importantly, from both surveys we see the same trend that
early-type or red galaxies have 2-3 times more massive halos than late-type or
blue galaxies at the same luminosity. 

\section{Summary}
\label{sect:summary}
We have presented a galaxy-galaxy lensing analysis of three fields from the
COMBO-17 survey. Although one of these fields is centered on a foreground
supercluster, we have shown that our measurements are not biased from the
presence of the supercluster. In addition, a cluster at $z=0.48$ behind the
supercluster was also found to have remarkably little influence. Despite the
limited area, the data set is unique at intermediate redshift in its extensive
redshift information that allows us to separate lens galaxies according to
their physical properties such as luminosity and rest-frame colours.

We fitted SISs and NFW profiles to the lens galaxies. In both cases we find
that red galaxies have dark matter halos about twice as massive as blue ones
when scaled to the same luminosity $L_\star=10^{10}h^{-2}L_{\sun}$. Within
$150h^{-1}~\mathrm{kpc}$ we obtain $M_\star=1.11^{+0.59}_{-0.54}\times
10^{12}h^{-1}M_{\sun}$ for blue galaxies and
$M_\star=2.26^{+0.64}_{-0.69}\times 10^{12}h^{-1}M_{\sun}$ for red galaxies
when modelled as SISs. Adopting NFW profiles we find virial masses
$M_\mathrm{vir}^*=3.9^{+3.3}_{-2.4}\times 10^{11}h^{-1}M_{\sun}$ for blue
galaxies and $M_\mathrm{vir}^*=7.1^{+7.1}_{-3.8}\times 10^{11}h^{-1}M_{\sun}$
for red galaxies. Note that the virial masses are defined as masses inside a
sphere with mean density equal to 200 times the mean density of the
universe. Changing this definition to the mass inside a sphere with mean
density equal to 200 times the critical density of the universe would lower
the virial masses by about 20\%. The mass estimates from the SIS model are
considerably larger than from the NFW profile although the virial radii exceed
the aperture adopted for the mass estimate of the SISs. However, it has
been shown by \citet{wright2000} that such behaviour is expected. 

For both models, we also fit the scaling between mass and luminosity. In the
SIS model we find approximately the same scaling for blue and red galaxies,
about $M_\star \propto L^{0.5}$ inside $150h^{-1}~\mathrm{kpc}$. For the NFW
model we find a similar scaling relation for blue galaxies, but
$M_\mathrm{vir}^*\propto L^{1.26}$ for red galaxies. The differences between
the two models and between the two subsamples for the NFW model might be due
to the different scales over which the fitted mass-luminosity relation
applies. For the SIS model it is always fitted over a fixed aperture of
$150h^{-1}~\mathrm{kpc}$ while for the NFW profile it only applies to the mass
within the virial radius that differs between blue and red galaxies.

Finally, we compared our results to those obtained from the RCS and
the SDSS. We pointed out that for such a comparison it is
indispensable to compare results from similar modellings. As far as is
possible we translated our measurements to the modellings of
\citet{hoekstra2004} for the RCS and \citet{guzik2002} for the SDSS and 
found remarkably consistent results. We think that this is remarkable,
at least because the modelling of \citet{guzik2002} is very different
from ours. A more exact comparison between COMBO-17 and the SDSS
adopting the same techniques is beyond the scope of this paper, although
such an investigation would be very valuable. The fact that both
surveys probe galaxies in different redshift ranges would in principle
allow measurement of the evolution of dark matter halos.

\begin{acknowledgements}
CW was supported by a PPARC Advanced Fellowship. MK acknowledges support by
the BMBF/DLR (project 50 OR 0106)), by the DFG under the project SCHN
342/3--1, and by the DFG-SFB 439 
\end{acknowledgements}

\bibliographystyle{aa}
\bibliography{bib}

\end{document}